\documentclass[hyper]{JHEP} 

\usepackage{epsfig}

\newcommand\fverb{\setbox\pippobox=\hbox\bgroup\verb}
\newcommand\fverbdo{\egroup\medskip\noindent%
			\fbox{\unhbox\pippobox}\ }
\newcommand\fverbit{\egroup\item[\fbox{\unhbox\pippobox}]}
\newbox\pippobox
\title{Particle Production on Half  S-brane}
\author{ J. Kluso\v{n}
\footnote{On leave from Masaryk University, Brno}\\
Institute of Theoretical Physics, University of Stockholm, SCFAB\\
SE- 106 91 Stockholm, Sweden \\
and \\
Institutionen f\"or teoretisk fysik\\
BOX 803, SE- 751 08 
Uppsala, Sweden \\
E-mail: \email{josef.kluson@teorfys.uu.se}}

\preprint{\hepth{0306002}}

\abstract{In this paper we will study  quantum 
field theory of fluctuation modes
 around  the rolling tachyon solution 
on  non-BPS D-brane effective action. 
The goal of this paper is to study 
particle production during the decay of non-BPS D-brane
and explore possible relation with minisuperspace calculation.
We find that the  number of particles 
produced on half S-brane 
exponentially  grows for large time which
suggests that linearised approximation breaks down
and also that backreaction of fluctuation field
on classical solution should be taken into account.}

\def\ket #1{\left|#1\right>}

\def\bk{\mathbf{k}}
\def\bx{\mathbf{x}}
\def\by{\mathbf{y}}
\def\bb{\mathbf{B}}

\def\bp{\mathbf{p}}

\def\mt{\mathcal{T}}
\def\ss{\sin \frac{\tau}{\sqrt{2}}}
\def\st{\sinh \frac{\tau}{\sqrt{2}}}
\def\st2{\sinh^2 \frac{\tau}{\sqrt{2}}}
\def\ss2{\sin^2 \frac{\tau}{\sqrt{2}}}
\begin{document}
\section{Introduction}\label{first}
A spacelike brane, or S-brane is almost the same as
ordinary D-brane except that one of its transverse
dimensions includes time. S-brane can be also seen
as time-dependent, soliton-like configurations
in a variety of field theories. In string
theory, the potential for the open string tachyon field
on the world-volume of unstable D-brane leads to 
S-branes \cite{Gutperle:2002ai} in a time-dependent
version of the construction of D-branes
as solitons of the open string tachyon.  These 
S-branes can be thought  as the creation
and subsequent decay of an unstable brane. This process
has recently attracted much attention
\cite{Gutperle:2002ai,Sen:2002an,
Sen:2002in,Sen:2002nu,Strominger:2002pc,Gutperle:2003xf,
Maloney:2003ck,Hashimoto:2003qx,Hashimoto:2002sk,
Kutasov:2003er,Lambert:2003zr,McGreevy:2003kb,
Moeller:2003gg,Gaiotto:2003rm,Lambert:2003zr,
Aref'eva:2003qu,Yang:2002nm,Sen:2002qa,Okuda:2002yd,
Mukhopadhyay:2002en,Moeller:2002vx,Sen:2002vv,
Minahan:2002if,Sugimoto:2002fp,Kluson:2003xu,Kluson:2002av,
Fujita:2003ex,Berkooz:2002je,Burgess:2002vu,Rey:2003xs,Rey:2003zj}
\footnote{For the most recent discussion of S-branes
and other time-dependent processes in string theory, 
see \cite{Gubser:2003vk,Giveon:2003gb,
Giveon:2003ge,Martinec:2003ka,Klebanov:2003km,
Constable:2003rc,McGreevy:2003ep,Demasure:2003av,
Kwon:2003qn}.}.
S-branes have been also extensively
studied in supergravity approach with potentially 
interesting cosmological applications
\cite{Kruczenski:2002ap,Chen:2002yq,
Leblond:2003db,Burgess:2003gg,Ohta:2003uw,
Gutperle:2003kc,Ohta:2003ie,Roy:2003nd,Ohta:2003pu,
Piao:2002nh,Shiu:2002cb,Buchel:2002tj,Buchel:2002kj,
Li:2002et,Mukohyama:2002cn,
Buchel:2003xa,Leblond:2003ac,McInnes:2003vu,Roy:2003ra,
Piao:2002vf,Guo:2003zf,Deger:2002ie,Deger:2003fz}.

Very nice world-sheet construction of S-branes was given
in the classical $g_s=0$ limit by A. Sen
\cite{Sen:2002an,Sen:2002in,Sen:2002nu} where
he introduced class of models in bosonic string theory
obtained by perturbing the flat space $c=26$ 
$CFT$ with the exactly marginal deformation
\begin{equation}\label{bound1}
S_{boud}=\lambda \int d\tau \cosh X^0(\tau) \ ,
\end{equation}
where $X^0$ is time coordinate, $t$ is a coordinate on the
world-sheet boundary and $\lambda$ is a free parameter
in the range $0\geq \lambda \leq \frac{1}{2}$
\footnote{We work in units $\alpha'=1$.}. This is family
of exact solutions of classical open string theory
whose space-time interpretation is that of an unstable brane
being created at a time $X^0 \sim -\tau$ and decaying
at a time $X^0 \sim \tau$ with $\tau=-\log (\sin (\pi\lambda))$. 
                                                             
As was shown  in  
\cite{Strominger:2002pc,Gutperle:2003xf,Maloney:2003ck} 
this time-dependent process of tachyon condensation 
 has many intriguing properties. 
In particular, it is known that in time-dependent backgrounds
there is in general no preferred vacuum and particle production
is unavoidable. In \cite{Maloney:2003ck}
the open string vacua on S-brane were studied and it
was shown that they have very mysterious properties. In 
particular,  for (\ref{bound1}) there is open string pair
production with a strength characterized by the Hagedorn
temperature $T_H=\frac{1}{4\pi}$ \cite{Strominger:2002pc,
Gutperle:2003xf}. As was argued 
\cite{Strominger:2002pc,Gutperle:2003xf,Maloney:2003ck} 
this temperature arises from the periodicity of the boundary
interaction (\ref{bound1}) in imaginary time. 

Recent intensive work on the dynamics of unstable D-branes
in string theory has led to an effective action 
for the open string tachyon $T$ and massless open string
modes $A_{\mu}$ (the gauge field on the D-brane) and
$Y^I$ (the scalar fields describing the location
of the D-brane in the transverse directions)
\cite{Sen:2002an,Sen:2002in,Sen:2003tm,
Hashimoto:2003qx,Hashimoto:2002sk,Hashimoto:2002sk,
Sen:2002qa,Kutasov:2003er,Lambert:2002hk,Lambert:2001fa,
Sen:2003bc,Sen:1999md,Garousi:2000tr,Bergshoeff:2000dq,
Kluson:2000iy,Lambert:2003zr}. As it was shown in
\cite{Kutasov:2003er,Lambert:2003zr}  the effective
action for non-BPS D-brane 
(\ref{nonact}), (\ref{Kutasovact})
reproduces several non-trivial aspects of open string
dynamics, for very nice recent discussion,
see \cite{Sen:2003tm,Kutasov:2003er,
Lambert:2002hk,Lambert:2001fa,Sen:2002qa,Sen:2003bc}.
These successes lead one to believe that the action 
(\ref{nonact}) captures some class of phenomena
in the full classical open string theory. 
As was argued in \cite{Kutasov:2003er} this action
should be thought of as a generalization of the DBI
action describing the gauge field $A_{\mu}$ and
scalars $Y^I$ on the brane 
\footnote{For review of DBI action, see
\cite{Tseytlin:1999dj}.}. 
On the other hand  we should be  more careful
when we speak about  the notion ``
low energy effective action'' in
case of unstable D-brane.
The fact that the tachyon has negative $m^2$ of the order
of the string scale, the notion of an effective action  which
is normally considered as a result of integration out
the heavy modes for describing the dynamics of light modes,
is somewhat unclear here. As was recently argued in 
\cite{Sen:2003bc}, the situation is even worse by the
fact that there are no physical states around the tachyon
vacuum and hence the usual method of deriving an effective
action-by comparing the S-matrix elements computed from string
theory with those computed from the effective 
action-does not work. The question of usefulness of the effective 
action for non-BPS D-brane was extensively discussed  in
\cite{Kutasov:2003er} where it was argued that 
its usefulness  can also be judged by comparing
the classical solution of the equation of motion derived
from the effective action with the classical solutions in
open string theory which are described by boundary conformal
field theory. In this respect the effective action 
constructed  in 
\cite{Sen:2003tm,Lambert:2003zr,Kutasov:2003er}
has had some remarkable success. 

Motivated by this 
success of effective action description 
of the open string phenomena on unstable
D-brane  and also very nice analysis
performed in \cite{Strominger:2002pc,Gutperle:2003xf,Maloney:2003ck} 
we asked the question whether there should
be some connection  between effective action description
of fluctuation fields around S-brane, or half S-brane,
 and minisuperspace description of
open strings on the same background. 
In other words, the goal of this paper
is to construct  quadratic  action for fluctuation
field  around
half S-brane solution on non-BPS D-brane.
 Since the classical solution describing
half S-brane  explicitly depends on time 
we can expect that metric and mass term in
the action for fluctuation field will be
function of  time as well. As is well known from
the study of quantum field theory in curved space-time
\footnote{For review, see \cite{Birrell:ix,Fulling:nb,Wald:yp}}
time-dependent background generally leads to 
the creation of particles during the time evolution.
Then we could expect that particles will be produced
during unstable D-brane decay as well. 
In order to study this process we will
 follow  \cite{Boyanovsky:1994me,Shtanov:1994ce,Martin:2002vn,
Kofman:1997yn,Greene:1997fu}.

Le us sketch the main idea of particle creation
on half S-brane. 
We start with non-BPS D-brane in unstable minimum
of the tachyon potential  in the asymptotic past 
$t\rightarrow -\infty$. This corresponds to the
situation
when the vacuum value of the tachyon is zero and
small fluctuation modes oscillate around it. 
Half S-brane solution then describes process when
the homogeneous tachyon field  starts to roll
from  unstable minimum of the potential ($T=0$) at
the time $t=-\infty$ 
to stable one ($T=\infty$) at asymptotic future
$t=\infty$. This rolling tachyon
 solution forms 
background for the fluctuation modes which is manifest-ally
time-dependent. It is important to 
know whether  these fluctuations, that were
small at the beginning of the rolling tachyon 
process,  remain small during time evolution
 or if they become large and hence they could affect
the classical evolution. Since large value of fluctuation
field can be interpreted as creation of large number
of particles we will be mainly interested in
the time evolution of this quantity. We 
find that number of particles created during
D-brane decay  exponentially grows at
asymptotic future. 
The similar behavior of the fluctuation
modes was observed in 
 \cite{Berkooz:2002je,Felder:2002sv,Frolov:2002rr},
where it was argued that the growing of fluctuation
modes quickly destabilize the linearised analysis
and hence backreaction of fluctuation field on
homogeneous solution should be taken into account. 
The result that we get in this paper also seems to
indicate that the linearised analysis of fluctuation
modes is only  suitable for the study of fluctuation at
the beginning of the rolling tachyon process. This is consistent
with the fact that only at the asymptotic past where 
the tachyon is in its unstable minimum 
  open string perturbative states exist  and 
one can define open string S-matrix. 
 It is also important to stress that
the analysis given in this paper  is restricted to the  
pure classical approximation where  we consider
the  string coupling
$g_s$ equal to zero. However it  was shown recently in
\cite{Lambert:2003zr,McGreevy:2003kb,
Gaiotto:2003rm,Klebanov:2003km,Constable:2003rc} that
the coupling between closed and open strings
in the  rolling tachyon process  plays fundamental
role which suggests that time-dependent 
D-brane decay is very 
complex problem and its study could answer some  fundamental
questions  considering nature of the string theory.

This paper is organized as follows. 
In the next section (\ref{second}) we briefly
review minisuperspace analysis of half S-brane
to show that particle production during
D-brane decay is natural process from the point
of view of quantum field theory on D-brane.
In section (\ref{third}) we introduce a
non-BPS Dp-brane effective action proposed
in \cite{Lambert:2003zr,Kutasov:2003er}.
Section (\ref{fourth}) is devoted to  study of
fluctuation modes above half S-brane solution
on non-BPS D-brane effective action.
In conclusion (\ref{fifth}) we outline 
the results obtained in this paper and give their
possible 
interpretation.

\section{Review of the minisuperspace approximation}\label{second}
In this section we present a short review of minisuperspace
approach to the study of S-brane dynamics, following
seminal papers  
\cite{Strominger:2002pc,Gutperle:2003xf,Maloney:2003ck}.

We wish to understand the dynamics of the open string world-sheet
theory with a time-dependent tachyon
\begin{equation}\label{action}
 S=-{1\over
4\pi}\int_{\Sigma_2} d^2\sigma \partial^a X^\mu \partial_a X_\mu + \int_{
\partial 
\Sigma_2} d\tau \ m^2(X^0) \ .
\end{equation}
For the open bosonic string $m^2=T$
where $T$ is the space-time tachyon, while for the open superstring
$m^2 \sim T^2$ after integrating out world-sheet fermions. We use
the symbol $m^2$ to denote the interaction because the coupling
(among other effects) impacts a mass to the open string states. We
consider three interesting classes described by the marginal
interactions
\begin{eqnarray}\label{bound}
 m^2_+(X^0) = \frac{\lambda}{2} e^{X^0} 
\\
m^2_-(X^0) = {\lambda \over 2} e^{-X^0} 
\\
 m^2_S(X^0)
= \lambda \cosh X^0  \ .
\end{eqnarray}
The first case $m_+^2$ describes the
process of brane decay, in which an unstable brane decays via
tachyon condensation. 
 The second case describes the time-reverse
process of brane creation, in which an unstable brane emerges from
the vacuum. The final case describes an S-brane, which is
the process of brane creation followed by brane decay. Brane decay
 can be thought of as the future (past) half of an
S-brane, i.e. as the limiting
case where the middle of the S-brane is pushed into the infinite past
(future).

In \cite{Strominger:2002pc,Gutperle:2003xf,Maloney:2003ck}
 this problem has been studied  using the minisuperspace
analysis  in which the effect of the interaction is simply to
give a time-dependent shift  to the masses of
all the open string states.  
In the minisuperspace approximation
only the zero-mode dependence of the interaction $m^2(X^0)$ is
considered.  In this case we can plug in the usual mode
solution for the free open string with oscillator number $N$ to
get an effective action for the zero modes 
\begin{equation}
 S = \int
d\tau \left(-\frac{1}{4}\dot{x}^\mu \dot{x}_\mu + (N-1) +
2m^2(x^0)\right) \  .
\end{equation}
 This is the action of a point particle with a
time-dependent mass. Here $x^\mu (\tau)$ is the zero mode part of
$X^{\mu} (\sigma,\tau)$, and the second term in  is an effective
contribution from the oscillators, including the usual normal
ordering constant. From upper action we can write down the Klein-Gordon
equation for the open string wave function $\phi(t,\bx)$,
\begin{equation}\label{minieq}
 \left(\partial^\mu\partial_\mu - 2{m^2(t)} -({N-1})
\right)\phi(t,\bx)=0, 
\end{equation}
 where $(t, \bx)$ are the space-time
coordinates corresponding to the world-sheet fields $(X^0,X^i)$.
 This is the equation of motion for a scalar field with
time-dependent mass. Quantum field theory of the scalar field
with time-dependent mass is similar with
with  the QFT in time-dependent  background 
 \cite{Birrell:ix,Fulling:nb,Wald:yp} 
hence we  we should make  few remarks about such field theories.
 Time translation invariance has been
broken, so energy is not conserved and there is no preferred set
of positive frequency modes. As a consequence 
this  leads in generally  to 
particle creation. The probability current
$j_\mu= i (\phi^*\partial_\mu \phi - \partial_\mu \phi^* \phi)$ is 
still
conserved, allowing us to define the Klein-Gordon inner product
\begin{equation}\label{kgcur}
\langle f|g\rangle = i \int_\Sigma d \Sigma^\mu
(f^*\partial_\mu g - \partial_\mu f^* g) \ , 
\end{equation}
 where $\Sigma$ is a spacelike
slice.  This norm does not depend on the choice of $\Sigma$ if $f$
and $g$ solve the wave equation. Normalized positive frequency
modes are chosen to have $\langle f|f\rangle=1$. Negative
frequency modes are complex conjugates of positive frequency
modes, with $\langle f^*| f^*\rangle=-1$. There is a set raising
and lowering operators associated to each choice of mode
decomposition -- these operators obey the usual oscillator algebra
if the corresponding modes are normalized with respect to (\ref{kgcur}).
We also define a vacuum state associated to each mode
decomposition -- it is the state annihilated by the corresponding
lowering operators.

To illustrate this idea we give an example of
the half S-brane corresponding to the decay of 
unstable  D-brane so that (\ref{minieq}) contains
following time-dependent mass term 
\begin{equation}
m^2_+(t)=\frac{\lambda}{2}e^t \ .
\end{equation}
Since in the past $t\rightarrow -\infty$ we have
ordinary  D-brane with  perturbative
open string spectrum it is natural to consider 
$\ket{in}$ vacuum as a vacuum with no particle present.
Expanding field in plane waves 
\begin{equation}
\phi(t,\bx)=e^{i\bp\bx}
u(t) 
\end{equation}
the wave equation becomes
\begin{equation}
(\partial_t^2+\lambda e^t+\omega^2)u=0 \ ,
\omega^2=p^2+N-1
\end{equation}
This is a form of Bessel's equation. It has normalized,
positive frequency solutions 
\begin{equation}
u^{in}=\lambda^{i\omega}\frac{\Gamma(1-2i\omega)}
{\sqrt{2\omega}}J_{-2i\omega}(2\sqrt{\lambda}
e^{t/2})
\end{equation}
where superscript $in$ and $out$  on a wave function 
denotes solutions that are purely positive frequency when
$t\rightarrow -\infty$ or $t\rightarrow \infty$. 
Using properties of Bessel's functions we can easily
show that $u^{in}$ approaches flat
positive frequency plane waves in the far past $t\rightarrow
-\infty$ 
\begin{equation}
u^{in}\sim \frac{1}{\sqrt{2\omega}}e^{-i\omega t} \ .
\end{equation}
We can also consider  the wave functions
\begin{equation}
u^{out}=\sqrt{\frac{\pi}{2}}
(ie^{2\pi\omega})^{-1/2}H^{(2)}_{-2i\omega}
(2\sqrt{\lambda}e^{t/2}) \ ,
\end{equation}
that are purely positive frequency in the far future
$t\rightarrow \infty$
\begin{equation}
u^{out}_+\sim \frac{\lambda^{-1/4}}{
\sqrt{2}}\exp \left\{-t/4-2i\sqrt{\lambda}e^{t/2}\right\} \ .
\end{equation}
Generally $u^{out}$ and $u^{in}$ are related by 
celebrated Bogolubov transformations
\begin{equation}\label{bol1}
u^{out}=Au^{in}+Bu^{in *} \ ,
\end{equation}
where coefficients $A,B$  can be determined from known
solutions $u^{in},u^{out}$. In particular, for modes
given above it can be shown that
\begin{equation}\label{bol}
A=e^{2\pi \omega+\pi i/2}B^*=
\sqrt{\omega}e^{\pi\omega-\pi i/4}\left(
\frac{\lambda^{-i\omega}}{\sinh 2\pi \omega 
\Gamma(1-2i\omega)}\right) \ .
\end{equation}
These coefficients obey unitarity relation
$|A|^2-|B|^2=1$.
The relation (\ref{bol1}) between in and out modes
implies the relation  between in and out creation
and annihilation operators
\begin{equation}
a_{in}=Aa_{out}+B^*a_{out}^{\dag} \ .
\end{equation}
From this relation we see that the condition $a_{in}\ket{in}=0$
implies that $\ket{in}$ is squeezed state
\cite{Maloney:2003ck}. Physically, this is the statement
that particles are produced during brane decay:
if we start in a state with no particles at $t\rightarrow
-\infty$, there will be many particles at
time $t\rightarrow \infty$. The density
 of particles of momentum $\bp$ is given
\begin{equation}
n_{\bp}=|\gamma|^2=e^{-4\pi\omega} \ . 
\end{equation}
We see that despite the fact that $\ket{in}$ is pure
state this is precisely the Boltzmann density of states at
temperature $T_H=1/4\pi$. In string units, $T_H$ is 
Hagedorn temperature. As was argued in
\cite{Strominger:2002pc,Gutperle:2003xf,Maloney:2003ck}
the fact that this temperature is so  high means
that $g_s$ corrections are likely
important even for $g_s\rightarrow 0$. 

The main conclusion which
we can get from this brief 
review of minisuperspace analysis
of S-branes  is that
 in the limit $g_s\rightarrow 0$ 
there is a  particle production during D-brane decay.
We would like to see the similar behavior from
the effective action point of view. We will discuss
 this problem in next sections. 

\section{Non-BPS D-brane effective action}\label{third}
 The dynamics of the tachyon field $T$ on
non-BPS D-brane  can be described by effective field theory where
the value of $T$ and its derivatives satisfy some conditions
\cite{Sen:2002qa,Kutasov:2003er,Lambert:2002hk}.
The effective action for real tachyon $T$ on
non-BPS Dp-brane in Type II string theory is expected
in the form
\cite{Sen:1999md,Garousi:2000tr,Bergshoeff:2000dq,
Kluson:2000iy}
\begin{equation}\label{nonact}
S=\int
d^{p+1}xL=-M_p\int d^{p+1}x
 \mathcal{V}(\mt)\sqrt{1+\eta^{\mu\nu}\partial_{\mu}
\mt\partial_{\nu}\mt} \ ,
\end{equation}
where $M_p$ is tension of non-BPS Dp-brane
\footnote{We are using the convention
where $\eta_{\mu\nu}
=diag(-1,1,\dots,1) \ , \mu, \nu=0,\dots, p \ ,
a,b=1,\dots,p \ $
 and the fundamental
string tension has been set equal to
 $(2\pi)^{-1} \ , \alpha'=1$.}.
Another form of the effective action for 
non-BPS D-brane has been obtained recently in
\cite{Kutasov:2003er}
\begin{equation}\label{Kutasovact}
S=\int d^{p+1}xL=-M_p\int d^{p+1}x
 V^2(T)\sqrt{1+\frac{T^2}{2}+
\eta^{\mu\nu}\partial_{\mu}T\partial_{\nu}T} \ ,
V^2(T)=\frac{1}{1+\frac{1}{2}T^2} \ ,
\end{equation}
where the tachyon field $T$ 
 given in (\ref{Kutasovact}) is related to $\mt $ in
 (\ref{nonact}) by the field redefinition
\begin{equation}\label{filedred}
\frac{T}{\sqrt{2}}=\sinh \frac{\mt}{\sqrt{2}} \ .
\end{equation}
We will concentrate in this paper on the action
(\ref{Kutasovact}) since it has 
the tachyon solution of equation of motion 
which has the same form as 
classical solution in open string theory that
in turn is  described by boundary conformal field theory
(BCFT). For that reason
 we believe that  field theory for fluctuation modes in
the effective action (\ref{Kutasovact}) 
could have some  relation to the quantum field theory
reviewed in the previous section that arises
as minisuperspace approximation of exact BCFT.

We would like also  study the behavior of massless
modes on  non-BPS D-brane effective action
during D-brane decay. For that reason we should
include them in (\ref{Kutasovact}). In order
to do this  we  use an equivalence between 
(\ref{Kutasovact}) and (\ref{nonact})
where the action 
(\ref{nonact}) containing 
massless modes is well known
 \cite{Sen:1999md,Garousi:2000tr,Bergshoeff:2000dq,
Kluson:2000iy}
\begin{equation}
S=-M_p\int d^{p+1}x \mathcal{V}(\mt)
\sqrt{-\det (\eta_{\mu\nu}
+\partial_{\mu}Y^I\partial_{\nu}
Y^J\delta_{IJ}+F_{\mu\nu}+\partial_{\mu}\mt\partial_{\nu}\mt}) \ ,
\end{equation}
where $A_{\mu}$ and $Y^I$ for $\mu \ , \nu=0,\dots,p \ ,
I=p+1,\dots,9$ are the gauge and the transverse scalar
fields on the world-volume of the non-BPS D-brane. 
If we expect that the field redefinition 
(\ref{filedred}) remains the same even  with nonzero
massless fields  and also that these massless modes
are the same in both forms of the action we
can write
\begin{equation}
T=\sqrt{2}\sinh\frac{\mt}{\sqrt{2}} \ ,
\mathcal{V}(\mt)=\frac{1}{\cosh\frac{\mt}{\sqrt{2}}}=
\frac{1}{\sqrt{1+\frac{T^2}{2}}} \ ,
\partial_{\mu}\mt=\frac{\partial_{\mu}T}{\sqrt{1+
\frac{T^2}{2}}} \ 
\end{equation}
and hence 
 we obtain generalized form of (\ref{Kutasovact})
\begin{equation}\label{Kutasovactm}
S=-M_p\int d^{p+1}x V(T)
\sqrt{-\det \mathbf{A}} \ , 
\end{equation}
where 
\begin{eqnarray}
\mathbf{A}_{\mu\nu}=
\eta_{\mu\nu}+\partial_{\mu}Y^I
\partial_{\nu}Y^J\delta_{IJ}+F_{\mu\nu}+
V^2(T)\partial_{\mu}T\partial_{\nu}T \ .
\nonumber \\ 
V(T)=\frac{1}{\sqrt{1+\frac{T^2}{2}}} \ , 
F_{\mu\nu}=\partial_{\mu}A_{\nu}-
\partial_{\nu}A_{\mu} \ .
\nonumber \\
\end{eqnarray}
Action (\ref{Kutasovactm}) will be useful
when we will discuss the massless mode
on half S-brane. For classical solutions
that depend on $T$ only the action 
 (\ref{Kutasovact}) is sufficient 
 so that the equation
of motion that arises from it is
\begin{equation}\label{eqKut}
-\frac{T}{\left(1+\frac{T^2}{2}
\right)^2}\sqrt{\bb}+\frac{T}{2(1+
\frac{T^2}{2})\sqrt{\bb}}-
\partial_{\mu}\left(\frac{\eta^{\mu\nu}
\partial_{\nu}T}{(1+\frac{T^2}{2})\sqrt{\bb}}\right)=0 \ ,
\end{equation}
where
\begin{equation}
\bb=1+\frac{T^2}{2}+\eta^{\mu\nu}\partial_{\mu}T
\partial_{\nu}T \ .
\end{equation}
Solution of (\ref{eqKut}) that is interpreted
as half S-brane is given by 
\begin{equation}\label{solhalf}
T_c=a\sqrt{2}e^{\frac{t}{\sqrt{2}}} \ ,
\end{equation}
where $a$ is some constant which
 can be set to $1$ by time
translation\cite{Lambert:2003zr} .

In the next section we present an  analysis
of the fluctuation modes above classical solution
(\ref{solhalf}).

\section{Analysis of fluctuations}\label{fourth}
In this section we will analyse the fluctuation
field around  the classical half S-brane solution
(\ref{solhalf}). To do this we write the
tachyon field $T$ that appears in
(\ref{Kutasovact}) as
\begin{equation}\label{Tgen}
T(t,\bx)=T_c(t)+\phi(t,\bx) \ ,
\end{equation}
where $T_c$ is given in (\ref{solhalf}) and
$\phi(t,\bx)$ is fluctuation field around $T_c$
 where we presume that
it is small with respect to the classical
solution. 
 
We insert  (\ref{Tgen}) into (\ref{Kutasovact})
and perform an expansion with respect to $\phi$
up to second order. As a result we obtain a quadratic
action for free massive scalar field $\phi$
 where  metric and mass term are
functions of classical solution $T_c$. More precisely,
let us write 
\begin{eqnarray}\label{actexp}
S=-\int dx L(T_c+\phi)=-\int dx L(T_c)-\nonumber \\
-\int dx \left(
\frac{\delta L(T_c)}{\delta T}\phi
+\frac{\delta L(T_c)}{\delta \partial_{\mu}T}
\partial_{\mu}\phi\right]-\nonumber \\
-\frac{1}{2}
\int dx \left(\frac{\delta^2 L(T_c)}{
\delta T\delta T}\phi^2+2
\frac{\delta^2 L(T_c)}{
\delta T\delta \partial_{\mu}T}\phi\partial_{\mu}\phi
+\frac{\delta^2 L(T_c)}{
\delta \partial_{\mu}T\delta \partial_{\nu}T}
\partial_{\mu}\phi\partial_{\nu}\phi\right)+\dots \ ,
\nonumber \\ 
\end{eqnarray}
where dots mean terms of higher order in $\phi$. 
Using  equation of motion and integration
by parts we can easily show
that the expression on the
second line in (\ref{actexp}) is equal to zero. 
The quadratic  effective
action for $\phi$  is then given on the third line
in   (\ref{actexp}). 
We can rewrite it in more familiar form using
\begin{eqnarray}
\int dx 
\frac{\delta^2 L(T_c)}{
\delta T\delta \partial_{\mu}T}\phi\partial_{\mu}\phi
=\int dx \partial_{\mu}\left[
\frac{\delta^2 L(T_c)}{
\delta T\delta \partial_{\mu}T}\phi^2\right]-\nonumber \\
-\int dx\partial_{\mu}\left[
\frac{\delta^2 L(T_c)}{
\delta T\delta \partial_{\mu}T}
\right]\phi^2-
\int dx
\frac{\delta^2 L(T_c)}{
\delta T\delta \partial_{\mu}T}
\phi\partial_{\mu}\phi \Rightarrow \nonumber \\
\Rightarrow
\int dx 
\frac{\delta^2 L(T_c)}{
\delta T\delta \partial_{\mu}T}\phi\partial_{\mu}\phi
=-\frac{1}{2}\int dx
\partial_{\mu}\left[
\frac{\delta^2 L(T_c)}{
\delta T\delta \partial_{\mu}T}
\right]\phi^2 \ ,  \nonumber \\
\end{eqnarray}
where as usual we do not carry about boundary
terms. Then 
we obtain  the action for fluctuation modes up
to second order in $\phi$ 
\begin{eqnarray}\label{fluctlin}
S=-\frac{1}{2}
\int dtd\bx \left[G^{\mu\nu}(T_c)\partial_{\mu}\phi
\partial_{\nu}\phi+M^2(T_c)\phi^2\right] \ , \nonumber \\
G^{\mu\nu}(T_c)=
\frac{\delta^2 L(T_c)}{
\delta \partial_{\mu}T\delta \partial_{\nu}T} \ , \ 
M^2(T_c)=\frac{\delta^2 L(T_c)}
{\delta T \delta T}-\partial_{\mu}\left[
\frac{\delta^2 L(T_c)}{\delta T\delta \partial_{\mu}T}\right] \ .
\nonumber \\
\end{eqnarray}
For half S-brane solution (\ref{solhalf})  
we obtain from  (\ref{Kutasovact}) 
 following metric and mass term that
appear in (\ref{fluctlin})
\begin{eqnarray}\label{met}
G^{00}=-\frac{M_pV^2}{\sqrt{\bb}}
-\frac{M_pV^2\dot{T}^2}
{
(\bb)^{3/2}}=-M_p  
\ , 
\nonumber \\
G^{ab}=\frac{M_pV^2\delta^{ab}}{\sqrt{\bb}}=
\frac{M_p\delta^{ab}}{1+e^{\sqrt{2}t}}
\equiv M_pG(t)\delta^{ab} \ ,\nonumber \\
M^2=-\frac{M_p}{2} \ ,
 \nonumber \\
\end{eqnarray}
where for (\ref{solhalf}) we have 
 $\bb=1+\frac{T^2}{2}-\dot{T}^2=
1$.  According to (\ref{met})
the  action for
fluctuation modes around  half S-brane solution
has the form 
\begin{equation}\label{halfSfluct}
S=-\frac{M_p}{2}\int dtd\bx
\left(-\partial_t \phi \partial_t\phi
+\frac{1}{1+e^{\sqrt{2}t}}\partial_i\phi
\partial^i\phi-\frac{1}{2}\phi^2\right) \ .
\end{equation}
This form of the action will be starting
point for the analysis of particle production
on half S-brane. 
 We will  show that (\ref{halfSfluct}) 
has  
similar form as  the action describing 
scalar  field  with time-dependent mass. Such
quantum field theories are well known, 
for very nice and detailed discussion,
see in particular \cite{Boyanovsky:1994me,
Shtanov:1994ce,Greene:1997fu,Kofman:1997yn}.

To begin  with, we should determine 
the conjugate momentum to $\phi$ 
and the Hamiltonian
that arise  from 
(\ref{halfSfluct})
\begin{eqnarray}\label{h}
\Pi(t,\bx)=\frac{\delta L}{\delta \dot{\phi}(t,\bx)}=
M_p\dot{\phi}(t,\bx) \ , \nonumber \\
\mathcal{H}(t,\bx)=\Pi(t,\bx)\dot{\phi}(t,\bx)-L=
\frac{1}{2}\left(\frac{\Pi^2}{M_p}+G^{ab}(t)\partial_a \phi\partial_b\phi
+M^2\phi^2\right) \ , \nonumber \\
H(t)=\int d\bx \mathcal{H}(t,\bx) \ . \nonumber \\
\end{eqnarray}
Since the effective metric and mass are functions of
time only   it
is natural to write operator $\phi(t,\bx)$ as
\cite{Shtanov:1994ce}
\begin{equation}\label{qp}
\phi(t,\bx)=\int d\bk Q_{\bk}(t)e^{i\bk\bx} \ ,
\Pi(t,\bx)=M_p\int d\bk P_{\bk}(t)e^{i\bk\bx} \ ,
\end{equation}
where  we have included
the normalization factor $\frac{1}{(2\pi)^p}$ into
the definition of $d\bk$. 
When we insert (\ref{qp}) into
the Hamiltonian we get
\begin{equation}
H(t)=\frac{M_p}{2}\int d\bk 
\left(P_{\bk}P_{-\bk}+
\left[G(t)k^2+\frac{M^2}{M_p}\right]Q_{\bk}Q_{-\bk}\right) \ .
\end{equation}
Condition  $\phi(t,\bx)=\phi^{\dag}(t,\bx)$ 
implies $Q^{\dag}_{\bk}=Q_{-\bk}$ 
and consequently  the Hamiltonian for the scalar field
$\phi$ 
can be seen as   collection of Hamiltonians $H_{\bk}$
of  quantum  oscillators with complex coordinates
$Q_{\bk}$ 
\begin{equation}\label{harm}
H=M_p\int d\bk H_{\bk} \ , 
H_{\bk}=\frac{1}{2}\left(P_{\bk}P_{\bk}^{\dag}+
\Omega^2_{\bk}(t)Q_{\bk}Q_{\bk}^{\dag}\right) 
\end{equation}
and with the time-dependent frequencies
\begin{equation}
\Omega_{\bk}^2(t)=G(t)k^2+\frac{M^2}{M_p} \ , 
\Omega_{\bk}=\Omega_{-\bk} \  \ ,k^2=\delta^{ab}k_ak_b \ .
\end{equation}
Canonical equal time  commutation relations for scalar field
\begin{eqnarray}
\left[\phi(t,\bx),\Pi(t,\by)\right]=i\delta (\bx-\by) \ , \nonumber \\
\left[\phi(t,\bx),\phi(t,\by)\right]=0 \ , \nonumber \\
\left[\Pi(t,\bx),\Pi(t,\by)\right]=0 \ , \nonumber \\
\end{eqnarray}
imply following commutation relations between
$Q_{\bk}  , P_{\bk}$ 
\begin{equation}\label{com}
[Q_{\bk},P_{\bk'}]=i\delta(\bk+\bk') \ ,
[Q_{\bk},Q_{\bk'}]=[P_{\bk},P_{\bk'}]=0 \ .
\end{equation}
In the following we will review the basic properties
of quantum oscillators with time-dependent frequencies. We
will mainly follow  
\cite{Shtanov:1994ce}. 

We would like to go to a time-dependent frame
in Hilbert space in which the Hamiltonian $H_{\bk}$
 is diagonal at every moment of
time. For that reason we define time-dependent
operators $a_{\bk}(t)$ and $a^{\dag}_{\bk}(t)$ by
\begin{eqnarray}
a_{\bk}=\frac{e^{i\int \Omega_{\bk} dt}}
{\sqrt{2\Omega_{\bk}}}(Q_{\bk}\Omega_{\bk}+iP_{\bk}) \ ,
\nonumber \\
a^{\dag}_{\bk}=
\frac{e^{-i\int \Omega_{\bk} dt}}
{\sqrt{2\Omega_{\bk}}}(Q^{\dag}_{\bk}
\Omega_{\bk}-iP^{\dag}_{\bk})=
\frac{e^{-i\int \Omega_{\bk} dt}}
{\sqrt{2\Omega_{\bk}}}(Q_{-\bk}\Omega_{\bk}-
iP_{-\bk}) \ .
\nonumber \\
\end{eqnarray}
These operators are Hermitian conjugate and have
the standard commutation relation of
creation-annihilation operators
\begin{equation}
[a_{\bk},a^{\dag}_{\bk'}]=\delta(\bk-\bk') \ .
\end{equation}
Using these operators one can write
\begin{eqnarray}
Q_{\bk}=\frac{1}{\sqrt{2\Omega_{\bk}}}
\left(e^{-i\int \Omega_{\bk} dt}a_{\bk}+
e^{i\int \Omega_{\bk} dt}a^{\dag}_{-\bk}\right) \ , 
\nonumber \\
P_{\bk}=\frac{\sqrt{2\Omega_{\bk}}}{2i}
\left(e^{-i\int \Omega_{\bk} dt}a_{\bk}-
e^{i\int \Omega_{\bk} dt}a^{\dag}_{-\bk}\right) \ 
\nonumber \\
\end{eqnarray}
so that  the Hamiltonian $H_{\bk}$ 
has the form 
\begin{equation}
H_{\bk}=\frac{\Omega_{\bk}}{2}\left(1
+a^{\dag}_{\bk}a_{\bk}+a^{\dag}_{-\bk}a_{-\bk}\right) \ .
\end{equation}
The orthonormal frame in Hilbert space which diagonalizes
the Hamiltonian $H_{\bk}$ is given by
acting creation operators $a^{\dag}_{\bk}$ on time
dependent vacuum state
$\ket{0_t}$ which is annihilated by the operator
$a_{\bk}(t)$. The operators $a_{\bk}(t) 
\ , a^{\dag}_{\bk}(t)$ 
obey  following equation of motion
\begin{eqnarray}\label{adageq}
\dot{a}_{\bk}=\frac{\dot{\Omega}_{\bk}}{2\Omega_{\bk}}
e^{2i\int \Omega_{\bk} dt}a^{\dag}_{-\bk} \ ,
\nonumber \\
\dot{a}^{\dag}_{\bk}=\frac{\dot{\Omega}_{\bk}}
{2\Omega_{\bk}}
e^{-2i\int \Omega_{\bk} dt}
a_{-\bk} \ . \nonumber \\
\end{eqnarray}
The solution of (\ref{adageq})  can be written
in terms of constant creation and annihilation
operators $a^{\dag}_{0,\bk} \ , a_{0,\bk}$ as follows
\begin{eqnarray}\label{bog}
a_{\bk}(t)=\alpha_{\bk}(t)
a_{0,\bk}+\beta^*_{\bk}(t)a^{\dag}_{0,-\bk} \ ,
\nonumber \\
a^{\dag}_{-\bk}(t)=\beta_{\bk}(t)a_{0,\bk}+
\alpha^*_{\bk}(t)a_{0,-\bk}^{\dag} \ .
\nonumber \\
\end{eqnarray}
If we insert (\ref{bog})  to (\ref{adageq})
we easily get
\begin{equation}
\dot{\alpha}_{\bk}=\frac{\dot{\Omega}_{\bk}}
{2\Omega_{\bk}}
e^{2i\int \Omega_{\bk} dt}\beta_{\bk}(t) \ , \ 
\dot{\beta}_{\bk}=
\frac{\dot{\Omega}_{\bk}}
{2\Omega_{\bk}}
e^{-2i\int \Omega_{\bk} dt}\alpha_{\bk}(t) \ .
\end{equation}
System (\ref{bog}) represents so named
Bogolubov transformation between two
pairs of creation-annihilation operators.
If the oscillator initially ($t=-\infty $) is in the vacuum
state then its state $\ket{0_0}$ is annihilated
by the operator $a_{0,\bk}$ and the initial condition
for functions $\alpha_{\bk}(t) \ , \beta_{\bk}(t)$ are
\begin{equation}
|\alpha_{\bk}(-\infty)|=1 \ , \beta_{\bk}(-\infty)=0 \ .
\end{equation}
At the moment $t$ it will not be in the vacuum
state $\ket{0_t}$ annihilated by the operator $a_{\bk}(t)$.
Recall that we are working in Heisenberg representation
where the state does not evolve. There is relation
between the states considered
\begin{equation}\label{inout}
\ket{0_0}=\prod_{\bk}
\frac{1}{\sqrt{|\alpha_{\bk}(t)|}}
\exp \left(\frac{\beta^*_{\bk}
(t)}{2\alpha^*_{\bk}(t)}
a^{\dag}_{\bk}(t)a^{\dag}_{-\bk}
(t)\right)\ket{0_t}
\ .
\end{equation}
At the moment $t$ the average number of
particles produced in the $\bk$-th
mode is
\begin{equation}\label{N}
N_{\bk}(t)=\left<0_0|a^{\dag}_{\bk}(t)
a_{\bk}(t)|0_0\right>=|\beta_{\bk}(t)|^2 \ .
\end{equation}
In Heisenberg representation the equations of motion
for $Q_{\bk}\ , P_{\bk}$ are
\begin{equation}
\dot{Q}_{\bk}=i[H_{\bk},Q_{\bk}]=P_{\bk} \ ,
\dot{P}_{\bk}=-\Omega^2_{\bk}(t)Q_{\bk} \ .
\end{equation}
Their  solutions can be written using
operators 
$a_{0,\bk} \ , a_{0,-\bk}^{\dag}$ as
follow
\begin{equation}
Q(t)=Q^{(-)}_{\bk}(t)a_{0,\bk}+
Q^{(+)}_{\bk}(t)a^{\dag}_{0,-\bk} \ ,
\end{equation}
where modes $Q^{(\pm)}_{\bk} \ ,
Q^{(-)*}_{\bk}(t)=Q^{(+)}_{-\bk}$
 obey differential equation
\begin{equation}\label{Qs}
\ddot{Q}^{(\pm)}_{\bk}+\Omega^2_{\bk}
(t)Q^{(\pm)}_{\bk}=0 \  .
\end{equation}
We can easily express $\beta_{\bk}(t)$ 
in terms of these modes
\begin{equation}
\beta_{\bk}(t)=\frac{1}{\sqrt{2\Omega_{\bk}}}
e^{-i\int \Omega_{\bk} dt}
\left(\Omega_{\bk} Q^{(-)}_{-\bk}-
i\dot{Q}^{(-)}_{-\bk}\right) \ 
\end{equation}
so that the  number $N_{\bk}(t)$ of particle
produced  is
equal to
\begin{equation}\label{Nt}
N_{\bk}(t)=\frac{\Omega_{\bk}}{2}\left(
\frac{\dot{Q}_{-\bk}^{(-)}\dot{Q}^{(+)}_{\bk}}
{\Omega^2_{\bk}}+Q^{(-)}_{-\bk}Q^{(+)}_{\bk}
\right)-\frac{1}{2}  \ .
\end{equation}
Now we apply the formalism reviewed
above to the case of fluctuation
field around
half S-brane solution (\ref{halfSfluct}).
Since the  number $N_{\bk}(t)$ of particle produced
 (\ref{Nt}) is function of $Q^{(\pm)}_{\bk}$
obeying  (\ref{Qs}), we must solve
this equation. 
In case when we do not need to know an exact solution 
the most efficient method to solve  (\ref{Qs}) is to use   
WKB approximation (For review of the
application of this method in  QFT in curved space-time,
see \cite{Birrell:ix}, for recent discussion, see
\cite{Martin:2002vn}.) In fact, we are mainly interested
in  estimation of  $N_{\bk}(t)$
 at asymptotic
future so that the WKB approximation 
can be applied.
Setting 
\begin{equation}
Q^{(-)}_{\bk}=\frac{1}{\sqrt{2W_{\bk}}}
\exp(-i\int^t dt W_{\bk}) \ 
\end{equation}
one can see that   the solution of equation 
(\ref{Qs})
is
equivalent to the condition
\begin{equation}\label{cond1}
W^2_{\bk}=\Omega_{\bk}^2
-\frac{1}{2}\left(\frac{\ddot{W}_{\bk}}
{W_{\bk}}-\frac{3}{2}\frac{\dot{W}_{\bk}^2}{W^2_{\bk}}
\right) \ . 
 \end{equation}
If following condition holds
\begin{equation}\label{conW}
\left|\frac{\mathcal{Q}_{\bk}}
{\Omega^2_{\bk}}\right|\ll 1 \ ,
\mathcal{Q}_{\bk}
\equiv\frac{1}{2}\left(
\frac{\ddot{\Omega}_{\bk}}
{\Omega_{\bk}}-\frac{3}{2}
\frac{\dot{\Omega}_{\bk}^2}{\Omega^2_{\bk}}
\right)
\ 
\end{equation}
we can write
\begin{equation}\label{sol1}
W_{\bk}(t)=\Omega_{\bk}(t) \ . 
\end{equation}
Let us apply  WKB approximation for equation (\ref{Qs}),
following very nice analysis given
in \cite{Martin:2002vn}. Firstly we observe
that there is a time $t_*$ when $\Omega_{\bk}^2$
vanishes
\begin{equation}
\Omega^2_{\bk}(t_*)=0 \Rightarrow
e^{\sqrt{2}t_*}=2\omega^2_{0,\bk}  \ , \omega^2_{0,\bk}
=
\bk^2-\frac{1}{2} \  ,
\end{equation}
i.e. $t_*$ is classical turning point. According
to \cite{Martin:2002vn} we define region I as
the region such that $\Omega^2_{\bk}>0$ and region
II the region where $\Omega^2_{\bk}<0$. 
 At some time $t$ deep in region
I the WKB approximation of $Q^{(-)}_{\bk}$ can be
written as
\begin{equation}\label{Qinft}
Q^{(-)}_{\bk}=\frac{A}{\sqrt{2\Omega_{\bk}}}
\exp(- i\int^t_{t_i} dt \Omega_{\bk}) \ ,
\end{equation}
where $t_i$ is the initial time at which the
normalization is performed.
We can also easily see that deep in region I the 
condition  (\ref{conW}) is valid.
Since for $t\rightarrow
-\infty$  the  half S-brane solution
(\ref{solhalf}) tends to zero 
corresponding  to the original non-BPS D-brane
it is natural to demand that for $t\rightarrow
-\infty$ 
$Q^{(-)}_{\bk}$ approaches the plane wave
mode
\begin{equation}
Q^{(-)}_{\bk}=\frac{1}{\sqrt{2\omega_{0,\bk}}}
\exp(-i\omega_{0,\bk}(t-t_i)) \  \ .
\end{equation}
  In fact,  deep in 
the region I we have
\begin{equation}
\Omega^2_{\bk}=\omega_{0,\bk}^2-\bk^2e^{\sqrt{2}t} \
\end{equation}
and hence  $Q^{(-)}_{\bk}$ is equal to 
\begin{equation}
Q^{(-)}_{\bk}=
\frac{A}{\sqrt{2\omega_{0,\bk}}}e^{-i\omega_{0,\bk}
(t-t_i)
+i\frac{\bk^2}{2\sqrt{2}\omega_{0,\bk}
}(e^{\sqrt{2}t}-e^{\sqrt{2}t_i})} \ .
\end{equation}
Comparing with (\ref{Qinft}) we 
immediately  obtain the value of normalization
constant $A$ equal to $1$. We must also stress that
we are considered the scalar modes with the initial
frequency  $\omega^2_{0,\bk}=\bk^2-\frac{1}{2}>0$
since these correspond to the standard fluctuation
modes on non-BPS D-brane in the past infinity. 
For modes that obey $\bk^2-\frac{1}{2}<0$
the analysis is much more complicated,  as was shown
in \cite{Felder:2000hj,Felder:2001kt}.  
 
In the region II the situation is different. In this
case $\Omega_{\bk}$ becomes complex. However this does not
prevent to use WKB approximation
 \cite{Martin:2002vn}. If we write $\Omega_{\bk}(t)$ as
$\Omega_{\bk}(t)=i|\Omega_{\bk}(t)|$ then the solution in
region II in WKB approximation is
\begin{equation}\label{Qreg2}
Q^{(-)}_{\bk}(t)=\frac{C_+}{|\Omega_{\bk}(t)|^{1/2}}
\exp \left[\int^t dt'|\Omega_{\bk}(t')|\right]
+\frac{C_-}{|\Omega_{\bk}(t)|^{1/2}}
\exp \left[-\int^t dt'|\Omega_{\bk}(t')|\right] \ .
\end{equation}
This solution should be matched on the solution
in the region I.  At the turning point $t=t_*$ 
the WKB approximation breaks down  and we
must use the usual WKB procedure \cite{Martin:2002vn}  
and approximate
the potential around the point $t_*$  by a straight
line such that
\begin{equation}\label{potaprox}
\Omega^2_{\bk}(t)\approx -\alpha(t-t_*) \ ,
\end{equation}
where
\begin{equation}
\alpha=-\frac{d\Omega^2_{\bk}(t_*)}{dt}=
\frac{2\sqrt{2}\omega^2_{0,\bk}(\omega^2_{0
,\bk}+\frac{1}{2})}
{1+2\omega^2_{0,\bk}}>0 \ .
\end{equation}
The solutions of the equation of motion
with such a potential are given in terms
of Airy functions of first and
second kinds
\begin{equation}\label{qair}
Q^{(-)}_k=B_1\mathrm{Ai}(s)+
B_2\mathrm{Bi}(s) \ , 
s\equiv \alpha^{1/3}(t-t_*) \ .
\end{equation}
Now we use the asymptotic behavior of the Airy 
functions to calculate the relation between 
$B_1 \ , B_2$ with $A$ on one hand and $C_{\pm}$
on the other. In region I, for a value of $t$ not
too far from $t_*$, (\ref{qair}) can be written
as 
\begin{eqnarray}\label{QI}
Q^{(-)}_{\bk,\mathrm{I}}(t)\approx
\frac{\alpha^{-\frac{1}{12}}}{2\sqrt{\pi}}
|t-t_*|^{-\frac{1}{4}}
\left[(B_1-iB_2)\exp\left(\frac{2i}{3}
\alpha^{\frac{1}{2}}|t-t_*|^{\frac{3}{2}}
+\frac{\pi i}{4}\right)\right.+\nonumber \\
\left. +
(B_2+iB_1)\exp
\left(-\frac{2i}{3}\alpha^{\frac{1}{2}}
|t-t_*|^{\frac{3}{2}}-\frac{\pi i}{4}
\right)\right] \ , \nonumber \\
\end{eqnarray}
while in region II, under the same conditions, the function
$Q^{(-)}_{\bk}(t)$ can be expressed as
\begin{equation}\label{ii}
Q^{(-)}_{\bk,\mathrm{II}}(t)\approx
\frac{\alpha^{-\frac{1}{12}}}{2\sqrt{2}}
(t-t_*)^{-\frac{1}{4}}
\left\{\frac{B_1}{2}\exp\left[-\frac{2}{3}
\alpha^{\frac{1}{2}}(t-t_*)^{\frac{3}{2}}\right]
+B_2\exp \left[\frac{2}{3}\alpha^{\frac{1}{2}}
(t-t_*)^{\frac{3}{2}}\right]\right\} \ .
\end{equation}
Now we evaluate in region I the solution
(\ref{Qinft}) for the potential (\ref{potaprox}).
The integral in the exponent can be written
as
\begin{equation}
\int_{t_i}^tdt'\Omega_{\bk}(t')=
\int_{t_i}^{t_*}dt'\Omega_{\bk}(t')+
\int_{t_*}^{t}dt'\Omega_{\bk}(t')
\equiv \Phi_{\bk}+\int_{t_*}^{t}dt'\Omega_{\bk}(t') \ .
\end{equation}
The frequency (\ref{potaprox}) is used in
the second integral only, assuming that
$t$ is not too far away from $t_*$. The
quantity $\Phi_{\bk}$ is just a number and its calculation
would require the knowledge of $\Omega_{\bk}$ in
the whole region I.  However as we will
see it does not enter in final result and therefore
we are not interested in its value. Then we find
\begin{equation}
Q^{(-)}_{\bk,\mathrm{I}}(t)=
\alpha^{-\frac{1}{4}}|t-t_*|^{-\frac{1}{4}}
\exp \left(\frac{2i}{3}\alpha^{\frac{1}{2}}
|t-t_*|^{\frac{3}{2}}+i\Phi\right)
\end{equation}
so that comparing with
(\ref{QI}) we find
\begin{equation}
B_1=iB_2 \ ,
B_2=\sqrt{\pi}\alpha^{-\frac{1}{6}}
e^{\Phi_{\bk}-\frac{\pi}{4}} \ .
\end{equation}
In the same way we can establish the link between
$C_+, C_-$ and $B_1,B_2$.  This can be
done in the same way as in the region I. 
If we define $\Psi_{\bk}$ as
\begin{equation}
\Psi_{\bk}
\equiv \int_{t_*}^{t_f}|\Omega_{\bk}(t')|dt'
\end{equation}
then (\ref{Qreg2}) reduces to the expression
\begin{eqnarray}
Q^{(-)}_{\bk,\mathrm{II}}(t)=\alpha^{-\frac{1}{4}}
(t-t_*)^{-\frac{1}{4}}\left\{
C_+e^{\Psi_{\bk}}\exp\left
[-\frac{2}{3}\alpha^{\frac{1}{2}}
(t-t_*)^{\frac{3}{2}}\right]+\right.\nonumber \\
\left.+ C_-e^{-\Psi_{\bk}}\exp
\left[\frac{2}{3}
\alpha^{\frac{1}{2}}(t-t_*)^{\frac{3}{2}}\right]
\right\} \ . \nonumber \\
\end{eqnarray}
By comparison with (\ref{ii}) we get
\begin{equation}
C_+=\frac{B_2}{2\sqrt{\pi}}
\alpha^{1/6}e^{-\Psi_{\bk}+i\pi/2} \ ,
C_-=\frac{B_2}{\sqrt{\pi}}\alpha^{\frac{1}{6}}
e^{\Psi_{\bk}} \ .
\end{equation}
The upshot of this analysis is that  generally 
both coefficients $C_+,C_-$ are nonzero 
when we match the solution in 
region II to the solution in region I  that corresponds
to the  plane
wave at asymptotic past $t\rightarrow -\infty$. For
 $t\gg t_*$ the leading term
in $Q^{(-)}_{\bk}$ 
is proportional to $C_-$ so that 
 we have
\begin{equation}
Q^{(-)}_{\bk}(t)=
\frac{B_2}{\sqrt{\pi}|2\Omega_{\bk}|^{\frac{1}{2}}}
\alpha^{\frac{1}{6}}e^{\Psi_{\bk}}
e^{-\int_t^{t_f} dt'
|\Omega_{\bk}(t')|}
\end{equation}
and consequently the   number
$N_{\bk}(t)$ of particle produced   for large $t$ is 
\begin{equation}
N_{\bk}(t)
=\frac{e^{2\Psi_{\bk}}}
{2}e^{-2\int_t^{t_f}dt'|\Omega_{\bk}(t')|}
-\frac{1}{2} \ . 
\end{equation}
In asymptotic future $t\rightarrow t_f$ and
 we get 
\begin{equation}
N_{\bk}(t)\sim e^{2\Psi_{\bk}}\sim e^{\sqrt{2}t_f} \ ,
\end{equation}
where we have used the fact that for $t\rightarrow \infty
$ $|\Omega_{\bk}|\rightarrow \frac{1}{\sqrt{2}}$. 
This result implies that 
during D-brane decay  infinite number of
particles is produced.  The result is
in agreement with
\cite{Berkooz:2002je,Felder:2002sv,Frolov:2002rr},
where similar  exponential instability of
fluctuation modes was found in
case of different non-BPS effective actions. 
It was then  argued there  that this
result implies that the approximation, where
we consider fluctuation modes as small with respect
to classical condensate and hence we can neglect
their backreaction, breaks 
down very early from the beginning of the time evolution.
We observe the same behavior in our approach as well.
More precisely, we can easily see that when we insert
 asymptotic form of fluctuation $\phi \sim e^{\frac{t}{\sqrt{2}}}$
to the action (\ref{halfSfluct}) that this
action  dominates over the 
action (\ref{Kutasovact})
evaluated
on half S-brane solution
\begin{equation}
S(T_c)=-M_p\int dt
\frac{1}{1+e^{\sqrt{2}t}} \ . 
\end{equation}
Then it follows  that 
the approximation when we
regard $\phi$ as small fluctuation modes without
any backreaction on classical solution breaks down
at late times. It would be nice to perform 
analysis of the fluctuation modes around half
S-brane solution when we take this backreaction
into account. We hope to return to this problem
in future. 

\subsection{Massless fluctuation on half S-brane}
In this section we extend our analysis to the case of
the massless fluctuations on half S-brane.
We start with the action
\begin{equation}\label{Kutasovactm1}
S=-M_p\int d^{p+1}xV(T)
\sqrt{-\det\left(\eta_{\mu\nu}+\partial_{\mu}Y^I
\partial_{\nu}Y^J\delta_{IJ}+F_{\mu\nu}+
V^2(T)\partial_{\mu}T\partial_{\nu}T\right)} \ .
\end{equation}
We will focus on scalar modes $Y^I$ only since
for gauge field $A_{\mu}$ the situation is
completely the same. 
When we perform an expansion of (\ref{Kutasovactm1})
around classical solution $T_c$  up to 
second order in $Y^I$ we get
\begin{eqnarray}
S=-M_p\int d^{p+1}xV(T_c)
\sqrt{-\det\left(\eta_{\mu\nu}
+V^2(T_c)\partial_{\mu}T_c\partial_{\nu}T_c+
\partial_{\mu}Y^I
\partial_{\nu}Y^J\delta_{IJ}
\right)}=\nonumber \\ 
=-M_p\int d^{p+1}xV(T_c)\sqrt{
-\det \mathcal{G}(T_c)}\sqrt{\det(\delta^{\mu}_{\nu}
+\mathcal{G}^{\mu\kappa}\partial_{\kappa}Y^I
\partial_{\nu}Y^J\delta_{IJ})}=\nonumber \\
=-M_p\int d^{p+1}xV(T_c)\sqrt{
-\det \mathcal{G}(T_c)}-\nonumber \\
-\frac{M_p}{2}\int d^{p+1}xV(T_c)\sqrt{
-\det \mathcal{G}(T_c)}
\mathcal{G}^{\mu\nu}(T_c)\partial_{\mu}Y^I
\partial_{\nu}Y^J\delta_{IJ} \ , \nonumber \\ 
\end{eqnarray}
where
\begin{equation}
\mathcal{G}(T_c)_{\mu\nu}=
\eta_{\mu\nu}+V^2(T_c)\partial_{\mu}T_c\partial_{\nu}T_c \ 
\end{equation}
which for half S-brane solution (\ref{solhalf})
 is equal to
\begin{equation}
\mathcal{G}(T_c)_{\mu\nu}=
\mathrm{diag}\left(-\frac{1+\frac{T^2_c}{2}-\dot{T}_c^2}{
1+\frac{T^2_c}{2}} \ , 1,\dots, 1\right) \ .
\end{equation}
 Consequently  the quadratic effective
 action for scalar  fluctuation
modes $Y^I$ around  half S-brane solution
has the form
\begin{eqnarray}\label{Yfluct}
S=-\frac{M_p}{2}\int d^{p+1}xV(T_c)\sqrt{
-\det \mathcal{G}(T_c)}
\mathcal{G}^{\mu\nu}(T_c)\partial_{\mu}Y^I
\partial_{\nu}Y^J\delta_{IJ}
=\nonumber \\
=-\frac{M_p}{2}\int d\bx dt
\left(-\partial_t Y^I\partial_tY_I+\frac{\delta^{ab}}{1+e^{\sqrt{2}t}}
\partial_a Y^I\partial_bY_I\right) \ .
\nonumber \\
\end{eqnarray}
Let us consider one  scalar mode,
say $Y^{p+1}\equiv Y$. 
 From (\ref{Yfluct})  we get conjugate  momentum 
to $Y$
\begin{equation}
\Pi(t,\bx)=\frac{\delta L}{\delta \dot{Y}(t,\bx)}=
M_p\dot{Y}(t,\bx) \ 
\end{equation}
so that  the Hamiltonian is
\begin{equation}
H=\frac{1}{2}\int d\bx 
\left(\Pi \dot{Y}-L\right)=
\frac{1}{2}\int d\bx
\left(\frac{\Pi^2}{M_p}+M_p G(t)^{ab}\partial_a Y\partial_b Y
\right) \ .
\end{equation}
If we take
take $Y=\int d\bk Y_{\bk}(t)e^{i\bk\bx}$ 
the Hamiltonian  reduces to the collection
of Hamiltonians $H_{\bk}$ with frequency $\Omega_{\bk}^2(t)$
\begin{equation}
H=\int d\bk H_{\bk} \ ,
H_{\bk}=\frac{1}{2}
\left(\frac{\Pi_{\bk}\Pi_{\bk}^{\dag}}{M_p}
+\Omega^2_{\bk}(t)Y_{\bk}
Y^{\dag}_{\bk}\right) \ ,
\end{equation}
where
\begin{equation}\label{omegamass}
\Omega^2_{\bk}(t)=\frac{M_p k^2}{1+e^{\sqrt{2}t}} \ .
\end{equation}
As we know from the previous section in order
to determine  number $N_{\bk}(t)$ 
of particle produced  at time $t$
 we should solve following equation
\begin{equation}\label{qmass}
\ddot{Q}^{(-)}_{\bk}+\Omega^2_{\bk}(t)Q^{(-)}_{\bk}=0 \ ,
\end{equation}
where $\Omega_{\bk}$ is given in (\ref{omegamass}). 
 As opposite to the case 
of tachyon fluctuation mode studied in previous
section from $\Omega_{\bk}$ given in  (\ref{omegamass})
we get
\begin{equation} 
\lim_{t\rightarrow \infty}
\frac{\dot{\Omega}_{\bk}}{\Omega_{\bk}}=
-\frac{1}{\sqrt{2}} \ ,
\lim_{t\rightarrow \infty}
\Omega_{\bk}(t)=0
\end{equation}
and hence WKB approximation cannot be used. 
However in order to estimate number of particles at
asymptotic future we does not need to know an exact solution
in the whole region, rather we could try to 
solve (\ref{qmass}) in asymptotic future. 
For large $t$ we can write 
\begin{equation}
\Omega_{\bk}^2(t)=k^2e^{-\sqrt{2}t}
\end{equation}
so that  (\ref{qmass}) is
\begin{equation}
\ddot{Q}^{(-)}_{\bk}+k^2e^{-\sqrt{2}t}
Q^{(-)}_{\bk}=0  \ .
\end{equation}
After substitution $m=\sqrt{2}ke^{-\frac{t}{\sqrt{2}}}$
we obtain following equation
\begin{equation}
m^2\frac{d^2Q^{(-)}_{\bk}}{d^2m}+
m\frac{dQ^{(-)}_{\bk}}{dm}+m^2Q^{(-)}_{\bk}=0 \ .
\end{equation}
This is Bessel's equation of the first kind 
with the  general solution
\begin{equation}
Q^{(-)}_{\bk}(m)=A_1 J_0(m)+A_2Y_0(m) \ .
\end{equation}
For $m\rightarrow 0$ we have
\begin{equation}
J_0(m)\sim 1-\frac{m^2}{4} \ ,
Y_0(m)\sim \ln m \ 
\end{equation}
and hence
\begin{equation}
Q^{(-)}_{\bk}(t)\sim A_1(1-\frac{k^2}{2}e^{-\sqrt{2}t})
+A_2 t \ .
\end{equation}
Let us consider for  a moment the situation when
$A_2=0$. Then the    number
$N_{\bk}$ of particle produced for large $t$ is equal to
\begin{equation}
N_{\bk}\sim \Omega_{\bk}
\left(\frac{A_1^2k^4}{2}
\frac{e^{-2\sqrt{2}t}}{\Omega_{\bk}^2}
+A_1^2\right)-\frac{1}{2}\sim
A_1^2\frac{e^{-2\sqrt{2}t}}{e^{-\frac{t}{\sqrt{2}}}}-\frac{1}{2}
\sim -\frac{1}{2} \ 
\end{equation}
which is  clearly nonphysical result since 
the  number of particles  cannot be 
 negative. For that reason any physical solution
must contain term proportional to $A_2$. 
However this term  dominates at large time so we
can write 
\begin{equation}
Q^{(-)}_{\bk} \sim A_2 t \ .
\end{equation}
For such a solution the 
number $N_{\bk}(t)$ of  massless particles
produced during D-brane decay is equal to 
\begin{equation}
N_{\bk}\sim\Omega_{\bk}
\left(\frac{A_2^2}{\Omega^2_{\bk}}+A_2t^2\right)
-\frac{1}{2}\sim \frac{A_2}{\Omega_{\bk}}\sim
e^{\frac{t}{\sqrt{2}}} \ .
\end{equation}
We see that the number of particles that are created
during the D-brane decay exponentially grows.  
This is consequence of the fact that for $t\rightarrow \infty$
the effective frequency $\Omega_{\bk}$ goes to zero.
Following
 \cite{Berkooz:2002je,Felder:2002sv,Frolov:2002rr}
we can interpret this exponential growth of particles
as breaking of linearised approximation
and their backreaction on classical solution
should be taken into account in the
process of D-brane decay.

\section{Conclusion}\label{fifth}
In this paper we have studied the particle
creation during  the time-dependent 
 process of non-BPS D-brane decay
(half S-brane) in the effective field
theory approximation. We hoped  
 that the behavior   of 
fluctuation modes around rolling tachyon solution 
on non-BPS D-brane could
reflect  the behavior of open string
modes as was described using 
 minisuperspace approach
\cite{Strominger:2002pc,Gutperle:2003xf,Maloney:2003ck}.
In particular, we have been mainly interested in
the question whether particle production occurs during
D-brane decay and whether 
 vacuum states of fluctuation modes  in asymptotic past and future
have similar properties as vacuum states introduced 
 in \cite{Strominger:2002pc,Gutperle:2003xf,Maloney:2003ck}.
While the particle production was unavoidable in
our effective action description as well since 
the action for fluctuation field around half S-brane solution
corresponds to the quantum field theory action
 in time-dependent background, 
we have find that linearised approximation is
not completely sufficient to reproduce results
given in \cite{Strominger:2002pc,Gutperle:2003xf,Maloney:2003ck}.
In particular we have shown that number of fluctuation
modes exponentially grows for large time  which implies 
that the linearised approximation breaks down.
Comparing the effective action for fluctuation field
given in this paper and the quantum field theory
action given in  
\cite{Strominger:2002pc,Gutperle:2003xf,Maloney:2003ck}
we immediately see  reason why our results are
so different. In 
\cite{Strominger:2002pc,Gutperle:2003xf,Maloney:2003ck}
effective frequency for given mode grows to infinity 
in the asymptotic future  and hence given 
mode exponentially decays. On the other hand, 
in the quadratic action for fluctuation field given
in previous sections  the time-dependent frequency
either approach constant value, as in the case of tachyon
fluctuation, or approaches  zero for the case of 
massless fluctuations. This behavior of effective
frequency  results into the  exponential growth of 
number of particles produced during D-brane decay.
 Similar exponential 
instability of fluctuations has been 
observed in
\cite{Berkooz:2002je,Felder:2002sv,Frolov:2002rr},
where this huge growth of  number of particles 
produced has
been interpreted as breaking of the linearised
approximation. 
To conclude we mean  that the  
linearised description of fluctuation modes around
half S-brane  is only efficient at
the beginning of the  rolling tachyon process. 
After some time it 
breaks down and we should  
take into account backreaction of fluctuation
field on the classical solution.

We must also stress that in this paper, following
\cite{Strominger:2002pc,Gutperle:2003xf,Maloney:2003ck},
we restrict  to the case $g_s=0$. 
However recent results given in
very interesting papers
\cite{Lambert:2003zr,McGreevy:2003kb,
Gaiotto:2003rm,Klebanov:2003km,Constable:2003rc}
show that in order to correctly describe
D-brane decay we should take into account 
interaction between closed and open strings
and hence to go beyond the free case  $g_s=0$. 

To conclude, we hope that  our modest results given
in this paper could be
helpful   for other study of non-BPS D-brane effective
action and time-dependent process of D-brane
decay. 
\\
\\
\\
{\bf Acknowledgment} I am graceful to
 J. Minahan and U. Danielsson
for useful conversations.
 This work is partly supported 
by EU contract HPRN-CT-2000-00122.

\end{document}